# 基于预训练语言模型和与知识蒸馏的轻量化序列推荐方法

李莉[1,2]　程明月[1,2]　刘芷町[1,2]　章浩[1,2]　刘淇[1,2]　陈恩红[1,2]

1. 中国科学技术大学，安徽 合肥 230026
2. 认知智能全国重点实验室，安徽 合肥 230000

**摘要：** 序列推荐通过用户历史行为记录建模用户兴趣，从而提供个性化推荐服务。现有的序列推荐算法主要采用神经网络模型对用户兴趣进行特征提取，取得了良好的推荐效果。但受限于推荐数据集的稀疏性，这些模型往往采用的是小型网络框架，泛化能力较弱。对此，近期提出了一系列基于大型预训练语言模型的序列推荐算法。但因推荐系统的实时性需求，如何在实际推荐场景中应用预训练语言模型实现快速推荐仍是一大挑战。为此，该文提出了一种基于预训练语言模型和知识蒸馏的序列推荐算法，核心是将预训练知识进行跨领域迁移并通过知识蒸馏实现轻量化推理。算法分为两个阶段：第一阶段使用预训练语言模型在推荐数据集上进行微调训练，将模型的预训练知识跨领域迁移到推荐中；第二阶段对训练好的语言模型进行知识蒸馏，将学习到的知识迁移到一个轻量级模型上。在多个推荐系统公开数据集上的实验表明，上述算法能在提升推荐精度的同时提供快速推荐服务。

**关键词：** 序列推荐；预训练语言模型；知识蒸馏
**中图分类号：** TP391　　　　**文献标识码：** A

# Pre-trained Language Model and Knowledge Distillation

# for Lightweight Sequential Recommendation

Li Li[1,2], Mingyue Cheng[1,2], Zhiding Liu[1,2], Hao Zhang[1,2], Qi Liu[1,2]and Enhong Chen[1,2]

(1. University of Science and Technology of China, Hefei , Anhui 230026, China ; 2. State Key Laboratory of Cognitive Intelligence, Hefei , Anhui 230000 ,China)

**Abstract:** Sequential recommendation models user interests based on historical behaviors to provide personalized recommendation. Previous sequential recommendation algorithms primarily employ neural networks to extract features of user interests, achieving good performance. However, due to the recommendation system datasets sparsity, these algorithms often employ small-scale network frameworks, resulting in weaker generalization capability. Recently, a series of sequential recommendation algorithms based on large pre-trained language models have been proposed. Nonetheless, given the real-time demands of recommendation systems, the challenge remains in applying pre-trained language models for rapid recommendations in real scenarios. To address this, we propose a sequential recommendation algorithm based on a pre-trained language model and knowledge distillation. The key of proposed algorithm is to transfer pre-trained knowledge across domains and achieve lightweight inference by knowledge distillation. The algorithm operates in two stages: in the first stage, we fine-tune the pre-trained language model on the recommendation dataset to transfer the pre-trained knowledge to the recommendation task; in the second stage, we distill the trained language model to transfer the learned knowledge to a lightweight model. Ex-tensive experiments on multiple public recommendation datasets show that the proposed algorithm enhances recommendation accuracy and provide timely recommendation services.

# 0 引言

现今，人们处于信息过载的大数据时代，从互联网日益增长的资源中获取到自己感兴趣的信息无异于大海捞针。推荐系统的出现正是为了解决这一问题，其根据用户历史行为记录对用户兴趣进行建模，提供个性化服务。在实际场景中，用户的历史行为通常存在时间上的顺序。序列推荐正是考虑到用户历史行为中的顺序依赖关系，通过建模连续的用户行为以及用户偏好随时间的动态演变，从而更容易地捕捉到用户的动态偏好，更准确地对用户下一个感兴趣的物品进行预测，提供更精准的个性化推荐[1]。

早期的序列推荐算法通常使用马尔科夫链和矩阵分解[2]来捕获用户行为序列的低阶顺序依赖。随后，一系列将用户行为序列建模为高阶马尔科夫链的扩展研究工作相继提出[3-4]。其中，基于深度神经网络的序列推荐算法能更好地捕捉到用户的动态偏好，取得了优异的推荐效果[5]。最初，循环神经网络(Recurrent Neural Network, RNN)及其变体[3,6-7]被引入序列推荐以捕获用户的长期偏好。尽管取得了不错的推荐效果，但由于 RNN 的网络结构难以进行并行计算[8-9]，导致算法的训练和推理速度都较慢，难以在大规模推荐场景中应用。随后，学者们开始探究除 RNN 之外的网络结构用于序列推荐，并且提出了大量研究工作[10-12]。

尽管许多深度学习模型提升了序列推荐效果，但受限于推荐数据集的稀疏性，这些模型通常采用小型网络架构，包括嵌入层、预测层以及几层中间隐藏层[14]，泛化性能较差。针对这一问题，近期有研究探索了基于大规模预训练语言模型(Pre-trained Language Model, PLM)的推荐范式。基于提示学习，Geng 等人[15]将各种推荐任务统一到一个框架中，通过预训练、个性化提示和预测来学习不同的任务。Zhang 等人[16]使用智能体

对现实推荐场景进行模拟，使模型的离线测试与在线表现更接近。然而，使用 PLM 会造成推理速度的下降，而推荐系统在实际应用中需要在用户浏览网页、观看视频或者在线购物时，及时快速地生成个性化推荐结果。因此，如何在实际推荐场景中应用 PLM 实现快速推荐仍是一大挑战。

为此，本文提出了一种结合预训练语言模型和知识蒸馏的序列推荐算法。其核心思想是将预训练模型从大规模语料库中获得的知识跨领域迁移到推荐场景中，并通过知识蒸馏将知识精炼萃取到一个小型模型中，从而在提升推荐效果的同时实现快速推荐。具体而言，受 Kao 等人[17]研究工作的启发，他们将在文本语料库上经过预训练的语言模型应用于生物分子序列的分类预测任务，并取得了令人满意的分类结果。这一实验结果表明预训练语言模型具有一定的通用性知识和强大的跨领域迁移能力。类似地，本文假设预训练语言模型学习到的知识可以迁移到推荐领域，并通过知识蒸馏得到一个轻量级模型用于个性化推荐。在公开数据集上的实验结果表明，该方法能在显著提升推荐精度的同时实现轻量化推理，满足推荐系统实时性的需求。

综上所述，本文的主要贡献包括：
1) 为了缓解数据稀疏性和实时性的问题，本文提出了一种结合预训练语言模型和知识蒸馏的序列推荐算法。
2) 在三个不同领域的公开数据集上进行了实验，实验表明本文提出的方法能在显著提升推荐精度的同时实现轻量化推理，满足推荐系统实时性的需求。

# 1 相关工作

本文旨在利用大规模预训练语言模型和知识蒸馏在提升推荐精度的同时实现轻量化推理。相关研究领域包括：序列推荐、预训练语言模型以及知识蒸馏。

## 1.1 序列推荐

序列推荐是一种通过用户与物品在时间序列

上的交互记录建模用户的动态偏好，以给用户推荐感兴趣物品的一种推荐系统范式。如图 1 所示，用户在某购物平台的浏览/购买记录以时间先后顺序依次排序，可见该用户正在选购移动通信相关产品，那么下一次给用户推荐"手机保护壳"等产品会更容易令用户购买。

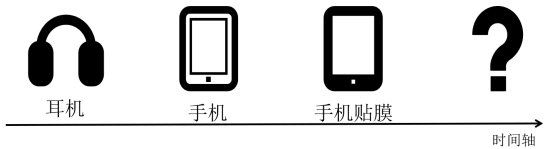

图 1 序列推荐

早期的序列推荐大部分是直接采用成熟的、经典的传统推荐算法，例如基于协同过滤和马尔科夫链的推荐算法。传统的协同过滤方法可以分为基于领域的方法和基于隐语义模型的方法。其中基于邻域的方法又可分为基于用户的协同过滤推荐[18]和基于物品的协同过滤推荐[19]。Sarwar 等人[20]基于奇异值分解计算用户和物品的潜在因子矩阵，利用潜在表示预测用户对特定物品的喜好程度。Liu 等人[21]采用主题模型(Latent Dirichlet Allocation, LDA)将兴趣作为隐藏维度，构建了用户-兴趣-物品的三层关系，实现了基于用户兴趣扩展的协同过滤增强。Steffen 等人[2]首次将马尔科夫链用于用户行为序列建模提出了因子分解算法 FPMC。He 等人[22]基于高阶马尔科夫链和物品相似度提出了 Fossil 算法，同时关注用户的前 L 个行为序列对当前预测的影响，可以兼顾用户的长期兴趣和短期兴趣。

随着深度学习技术的迅速发展，基于深度学习的序列推荐算法大量涌现，逐渐成为主流。2015 年，Hidasi 等人[3]基于循环神经网络创新性地提出了 GRU4Rec 模型，证实了深度神经网络用于序列推荐的可行性。之后，Tang 等人[8]使用卷积神经网络中不同的卷积滤波器学习用户的行为特征，对用户兴趣进行建模。Yuan 等人[9]提出了 NextItNet 模型，利用空洞卷积对用户行为序列进行建模。随着注意力机制(Attention Mechanism)[23]的提出，Kang 等人[4]基于Transformer 模型[24]提出了 SASRec 模型，能够借助注意力机制更加关注用户近期的交互记录。基于双向注意力模型 BERT[24]，Sun 等人[10]提出

了 BERT4Rec 模型，通过捕获用户和物品之间的上下文信息来提高推荐精度。此外，Wu 等人[26]首次将图神经网络应用于序列推荐提出了SR-GNN 模型。Cheng 等人[27]利用对比学习提出了一种自监督方法用于增强用户表征以提高序列推荐的效果。

## 1.2 预训练语言模型

预训练语言模型是一种通过在大规模无标注文本数据上进行预训练来学习语言特征以及常识知识的深度学习模型。通过无监督学习方式，PLM 能在预训练阶段学习文本的语法、语义和上下文信息，然后在特定的下游任务上进行微调，从而实现各种语言处理任务的高效处理。常见的预训练语言模型包括 BERT[25]、GPT[28]、T5[29]。

预训练语言模型具有强大的跨领域迁移能力。Kao 等人[17]认为生物分子序列与自然语言序列存在共性知识，将在文本数据上经过预训练的语言模型 BERT 应用于蛋白质、DNA 等生物分子序列的分类预测任务，并获得了较好的分类效果。这一实验结果表明了 PLM 可以高效地适应不同学科的数据，具有一定的通用性知识以及很好的泛化性。基于此，本文进一步假设用户交互序列与自然语言序列间也存在着共性知识，PLM 也可以适应用户行为序列数据。

近期 PLM 也被引入推荐领域，提出了新的推荐范式。基于提示学习，Geng 等人[15]为各种推荐任务设计了一个统一的框架 P5，通过预训练、个性化提示和预测来学习不同的任务。Gao 等人[30]提出了一种基于会话的推荐范式Chat-Rec，将用户信息和历史交互转换为提示通过 PLM 构建会话系统，使得推荐过程更具可解释性。Zhang 等人[16]使用智能体对现实推荐场景进行模拟，使模型的离线测试与在线表现更接近，提升模型的在线性能。

## 1.3 知识蒸馏

知识蒸馏(Knowledge Distillation)是一种模型压缩技术，其目的是通过将大型复杂模型的知识迁移到小型简单模型上，以提升小型模型的性

能。在知识蒸馏过程中，通常将大型复杂模型作为"教师"模型，小型简单模型作为"学生"模型。通过预先训练好的教师模型提供知识，以轻微的性能损失为代价将复杂教师网络的知识蒸馏到简单的学生网络中，最终训练出一个轻量级网络以逼近复杂网络的性能。

早期，Buciluǎ 等人[31]首次提出模型压缩的思想，强调知识的迁移而不是权重的迁移，它通过学习大型复杂模型的近似特征来生成轻量级网络模型[32]。Ba 等人[33]利用 L2 损失函数通过训练学生网络来模拟教师网络的逻辑输出单元。Li 等人[34]利用神经网络输出分布特征，最小化教师网络和学生网络输出之间的 KL 散度。

早期的工作都利用了教师网络的逻辑单元和类概率来进行模型压缩，然而使用 Softmax 函数处理的类概率容易丢失负标签处的概率信息。为了解决这一问题，Hinton 等人[35]通过引入软目标，结合硬目标即正确的数据标签，使用两个不同的损失函数加权平均来训练学生模型。具体流程如图 2 所示。

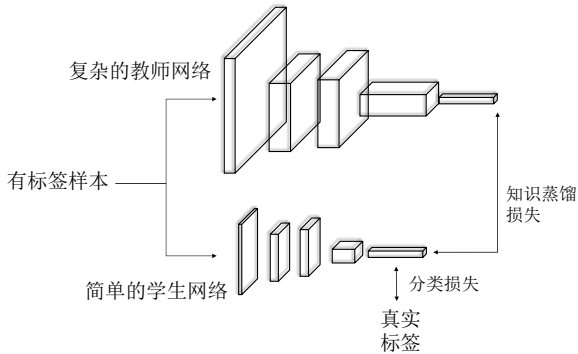

图 2 知识蒸馏流程

## 2 基于预训练语言模型和知识蒸馏的轻量化序列推荐

本文旨在将预训练语言模型从大规模语料库中获得的知识跨领域迁移到推荐场景中，并通过知识蒸馏精炼萃取大模型中的海量知识到小模型上，从而在提升推荐精度的同时实现快速推荐。基于上述思想，本文提出了一种两阶段的序列推荐算法，本节将以 BERT 作为预训练语言模型样例介绍算法流程。第一阶段利用在文本数据上经过预训练的 BERT 模型，使用与掩码语言模型(Masked Language Model, MLM)一致的训练范式，对用户交互序列中的部分进行遮蔽(Mask)，将被遮蔽的物品作为模型的自监督信号，从而使模型能够学习用户交互记录中的双向序列信息；第二阶段基于已经训练好的 BERT 模型，将其学习到的知识蒸馏到一个小型简单模型上，得到一个轻量化推理模型。

### 2.1 第一阶段：基于预训练语言模型的序列推荐

#### 2.1.1 BERT 模型框架

BERT 由多层双向的 Transformer 编码器堆叠而成，具有双向的编码能力和强大的序列建模能力。模型的结构大致上可分为特征嵌入层，Transformer 编码器的堆叠，预测层。

特征嵌入层包括三个部分，分别是词嵌入(Token Embedding)、段落嵌入(Segment Embedding)、位置嵌入(Position Embedding)，三种嵌入分别对词的含义、词的段落位置、词的位置进行特征表征，三种不同的嵌入求和为 BERT 的输入向量。

Transformer 编码器包括多头注意力层、残差连接、层归一化以及前馈网络层，其中多头注意力层最为关键。注意力权重采用缩放点积进行计算。

$$Attention(\boldsymbol{Q}, \boldsymbol{K}, \boldsymbol{V}) = softmax(\frac{\boldsymbol{QK}^T}{\sqrt{d_k}})\boldsymbol{V} \quad (1)$$

其中，$\boldsymbol{Q}$、$\boldsymbol{K}$、$\boldsymbol{V}$ 为向量矩阵，$d_k$为向量维度。注意力权重代表了序列中上下文信息对当前词的影响程度。多头注意力机制中，在逻辑上将输入向量分为多个部分，并在每个部分上进行独立的注意力计算，这些部分被称为"头"，每个头可以关注到不同的序列特征。

$$MH(\boldsymbol{H}^l) = [head_1; head_1; \cdots; head_h]\boldsymbol{W}^O \quad (2)$$

$$head_i = Attention(\boldsymbol{H}^l\boldsymbol{W}_i^Q, \boldsymbol{H}^l\boldsymbol{W}_i^Q, \boldsymbol{H}^l\boldsymbol{W}_i^Q) \quad (3)$$

Transformer 在多头注意力之上，又进行了残差连接、层归一化和 Dropout 操作，都是为了缓解深层模型的过拟合以及梯度消失或爆炸的问题，以实现模型的堆叠。注意力层采用线性投影的方式，为了使模型具备非线性拟合能力最后加入了两层前馈网络。

#### 2.1.2 数据预处理

BERT 作为语言模型，使用的数据为文本语

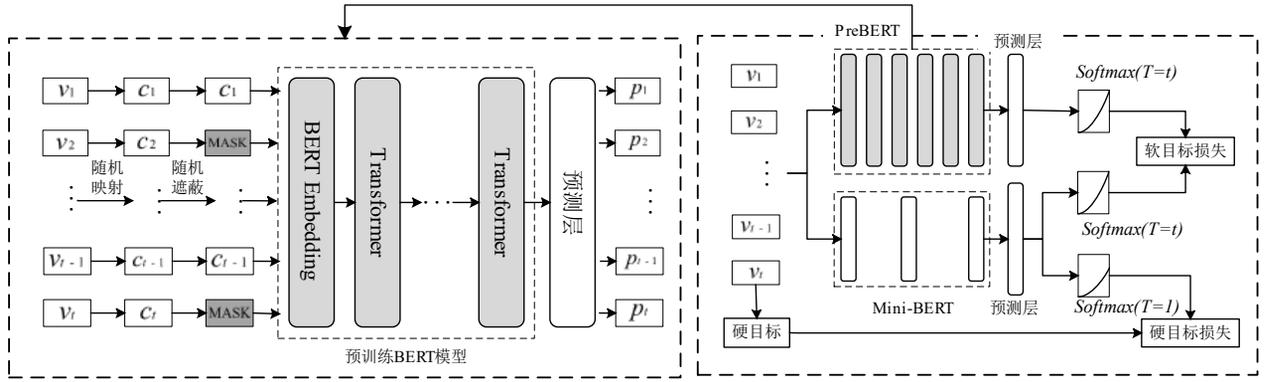

图 3  算法框架

使用的数据为文本语句，将分词器分词得到的词 (Token)序列作为模型输入，每个词对应唯一的 ID。序列推荐中也采用 ID 范式，每个物品对应唯一的一个 ID。参考 Kao 等人[15]的工作，其将不同的氨基酸随机映射到不同的 Token，本文将不同的物品一一随机映射到不同的 Token。如此，这一映射实现了从物品到文本的跨越，使预训练的 BERT 可以用于推荐领域中。

在数据预处理上，本工作参考了大多数工作的处理方式[3-4]，略去了交互记录少于 5 的物品和用户。为了保证训练的可行性，需要所有用户的交互序列长度保持一致。若序列长度大于 $n$，则截取最近交互的 $n$ 个物品，若序列长度小于 $n$，则在序列前填补上特殊的 ID，使序列长度达到 $n$。

### 2.1.3 基于掩码的模型训练

传统的单向序列推荐模型通常通过预测输入序列中每个位置的下一个物品来训练模型，输入序列[$v_1, v_2, \cdots, v_t$]，预测目标[$v_2, v_3, \cdots, v_{t+1}$]，其中 $v_i$ 是用户在 $i$ 时刻交互的物品。但在双向模型中，模型能"看到"序列中的每个物品，如以下一个物品为预测目标，网络将学不到任何有用的信息。一个简单的解决方案是为序列中的每个位置创建子序列，如如([$v_1$], $v_2$)和([$v_1, v_2$], $v_3$)等，利用双向模型对每个子序列进行目标物品的预测。然而，这种方法非常消耗时间和资源。为了更高效地训练 BERT，本文采用"完形填空"即 MLM 的训练方式，通过对序列中的部分物品进行遮蔽，然后预测遮蔽位置原始内容来训练网络。

具体的遮蔽方式是以概率 $\rho$ 随机对序列中的物品进行遮蔽，其中有 80%的概率用遮蔽标记 [MASK]进行替换，有 10%的概率用其他物品进行替换，有 10%的概率保持不变。由于每次输入时都会随机改变遮蔽位置，因此采用这种遮蔽方式可以产生更多不同的输入用于训练模型，保证了有足够的数据支撑 BERT 的训练。该阶段的训练框架如图 3 左半部分所示，把训练得到的模型称为 PreBERT。

通过预测遮蔽位置的原始内容，结合交叉熵函数，损失函数可以写成：

$$\mathcal{L} = \frac{1}{|S_u^m|} \sum_{v_m \in S_u^m} -\log P(v_m = v_m^* | S_u) \quad (4)$$

其中 $S_u$ 是用户行为序列被部分遮蔽后的序列，$S_u^m$ 是序列中被遮蔽物品，$v_m^*$ 是遮蔽物品 $v_m$ 的原始值。

## 2.2 第二阶段：知识蒸馏

### 2.2.1 带温度系数的 Softmax 函数

Softmax 函数是一种激活函数，通常用作神经网络最后一层的输出函数，输出一个概率分布。其公式如下：

$$softmax(z_i) = \frac{e^{z_i}}{\sum_j e^{z_j}} \quad (5)$$

其中 $\mathbf{z} = (z_1, z_2, \cdots, z_n)$ 是输入向量，Softmax 将其归一化为概率分布，使每个元素的范围为[0,1]，所有元素总和等于 1。

具有温度系数 $T$ 的 Softmax 函数被定义为：

$$softmax(z_i, T) = \frac{e^{z_i/T}}{\sum_j e^{z_j/T}} \quad (6)$$

温度系数 $T$ 的引入可以控制输出概率分布的随机性和多样性。具体来说，当 $T = 1$ 时即为标准的 Softmax 函数；当 $T > 1$ 时，元素之间的差异会减小，使输出的概率分布更加平滑，随机性增大；当 $T < 1$ 时，元素之间的差异会变大，输出的概率分布更加尖锐，确定性增大。

### 2.2.2 蒸馏过程

参考 Hinton 等人[35]的蒸馏方法，本文使用第一阶段中训练好的 PreBERT 模型作为教师网络，为保持训练范式一致，使用两层的 BERT4Rec 模型[10]作为学生网络，通过引入带有温度系数的 Softmax 函数，将要预测的真实用户交互序列作为硬目标，PreBERT 预测层输出的概率分布作为软目标，利用两个目标相结合将教师网络学习到的知识迁移到学生网络上，旨在获得一个轻量化的模型 Mini-BERT，实现轻量化推理。Softmax 的温度系数 $T$ 也被称为蒸馏的温度。

BERT4Rec 模型与 BERT 类似，也是由 Transformer 编码器堆叠而成，但隐藏层向量维度相比于 BERT 小得多，本文使用两层 Transformer 编码器堆叠而成的 BERT4Rec 模型。由于 BERT4Rec 模型也采用双向注意力机制，因此训练范式与前文所述相同，也是采用 MLM 的训练方式。在测试阶段，只需在序列最后添加一个特殊标记[MASK]，即可对用户下一个交互的物品进行预测。

通过引入带有温度系数 $T$ 的 Softmax 函数，升高温度可以使输出的概率分布更平滑，更具有随机性。将 PreBERT 模型逻辑单元输出的得分经过带温度的 Softmax 函数后得到的概率分布作为软目标，此时软目标相比于物品真实标签的硬目标可以提供更多的信息。同时软目标的加入相当于增添了一个正则项，可以增强模型的泛化能力。使用交叉熵函数度量两个概率分布之间的差异，软目标和硬目标的损失函数可以写成以下形式：

$$\mathcal{L}_{soft} = H(softmax(\mathbf{z}_s, t), softmax(\mathbf{z}_t, t)) \quad (7)$$

$$\mathcal{L}_{hard} = H(softmax(\mathbf{z}_s, t), \mathbf{v}^*) \quad (8)$$

$$H(\mathbf{p}, \mathbf{q}) = -\sum_i p(x_i) q(x_i) \quad (9)$$

其中 $\mathbf{z}_s$、$\mathbf{z}_t$ 分别是学生网络和教师网络的逻辑单元在被遮蔽位置的输出，$\mathbf{v}^*$ 为待预测物品序列的真实值。

为了结合软目标和硬目标，将软目标和硬目标的损失函数进行加权平均得到最终的损失函数。由于软目标产生的梯度大小前带有系数 $\frac{1}{T^2}$，因此加权平均时需在软目标损失函数前乘以 $T^2$，以确保在更改了蒸馏的温度后，硬目标和软目标的相对贡献大致保持不变。总损失函数如下：

$$\mathcal{L} = \alpha \mathcal{L}_{hard} + (1 - \alpha) T^2 \mathcal{L}_{soft} \quad (10)$$

其中 $\alpha \in [0,1]$ 为加权系数，$\alpha$ 越小，硬目标即真实标签的影响越小，软目标即 PreBERT 输出的概率分布影响越大。

本阶段具体流程图如图 3 右半部分所示。

## 3 实验

### 3.1 实验数据

本文在以下三个不同领域的公开推荐数据集上进行探究实验。

**ML-1M**：MovieLens 即 ML-1M 数据集是由 GroupLens 项目组从 MovieLens 网站收集整理的一个电影评分数据集，含有 6040 名用户和 3706 部电影，涵盖了大约 100 万条用户对电影的评分。

**Steam**：Steam 数据集记录了游戏购买平台 Steam 上用户对众多游戏的评论信息等，包含了大约 256 万名用户对其中 15474 个游戏的 779 万条评论。

**YooChoose-Buys**：YooChoose 数据集收集了欧洲在线零售商平台上用户的点击/购买记录，本文只使用了用户购买记录数据集，其中包含了大约 2 万个商品、50 万个用户和 115 万条用户购买记录。

在数据处理上，本工作参照大多数工作的处理方法，为了保证数据集的质量，对数据集中交互记录少于 5 的用户和物品进行忽略，经过筛选后的数据集的统计信息如表 1 所示。对于所有数据集，本工作将所有数字评分或是否有评论均看作隐式反馈。

表 1 数据集统计信息

| 数据集 | 用户数 | 物品数 | 交互数 | 用户平均交互数 | 物品平均交互数 |
|---|---|---|---|---|---|
| ML-1M | 6040 | 3416 | 999611 | 165.50 | 292.63 |
| Steam | 334542 | 13046 | 4212380 | 12.59 | 322.89 |
| YooChoose | 45776 | 8490 | 311160 | 6.80 | 36.65 |

在数据集划分上，本文采用留一法对用户交互序列进行划分，将用户的最后一个交互物品作为测试集，倒数第二个交互物品作为验证集，其余作为训练集。

## 3.2 对比方法

为了证明本工作提出的算法的有效性，本文将对比几种经典的序列推荐算法，包括传统的序列推荐算法和基于深度学习的序列推荐算法。

**BPR-MF**[36]：一种传统的序列推荐算法，利用 BPR(Bayesian Personalized Ranking)损失函数对用户-物品的潜在因子矩阵进行优化。

**GRU4Rec**[3]：使用门控循环单元(Gate Recurrent Unit, GRU)对用户行为序列进行建模。

**Caser**[8]：使用卷积神经网络利用水平卷积层和垂直卷积层对用户行为序列进行高阶马尔科夫链建模。

**SASRec**[4]：使用从左到右的单向 Transformer 模型来捕捉用户的行为序列。

**BERT4Rec**[10]：使用基于 BERT 的双向注意力机制对用户的行为序列进行建模。

## 3.3 实验设置

### 3.3.1 第一阶段：预训练语言模型参数设置

本工作中使用的预训练语言模型为 BERT 模型，其中预训练参数权重为在 BookCorpus 数据集上经过训练的"BERT-base-cased"模型。模型由 12 层 Transformer 编码器堆叠而成，隐藏维度为 768，多头注意力个数为 12，Dropout 比率为 0.1，训练时在 ML-1M 和 Steam 数据集上使用的遮掩比率为 0.55，在 YooChoose 数据集上使用的遮掩比率为 0.35，学习率设置为 2e-5。

### 3.3.2 第二阶段：蒸馏阶段参数设置

蒸馏阶段使用的学生网络模型为 BERT4Rec 模型，由 2 层 Transformer 编码器堆叠而成，隐藏维度为 256，多头注意力个数为 4，Dropout 比率为 0.1，学习率设置为 1e-4，其他参数(遮掩比率$\rho$，损失函数系数$\alpha$，蒸馏温度$T$)在不同数据集上设置如下表 2 所示。

表 2 蒸馏阶段参数设置

| 数据集 | $\rho$ | $\alpha$ | $T$ |
|---|---|---|---|
| ML-1M | 0.35 | 0.5 | 1.5 |
| Steam | 0.45 | 0.7 | 1.5 |
| YooChoose | 0.35 | 0.5 | 1.3 |

### 3.3.3 其他对比方法参数设置

所有对比模型均由两层基本结构堆叠而成，隐藏维度均为 256，不同的参数如下表 3 所示。

表 3 其他对比方法参数设置

| 参数 | BPR-MF | GRU4Rec | Caser | SASRec | BERT4Rec |
|---|---|---|---|---|---|
| 多头注意力个数 | – | – | – | 4 | 4 |
| Dropout | 0 | 0.2 | 0.2 | 0.5 | 0.1 |
| 学习率 | 1e-3 | 1e-3 | 1e-3 | 1e-3 | 1e-4 |

所有模型输入的序列最大长度均为 50，批处理大小均为 32，使用的优化器均为 Adam 优化器，优化器的参数$\beta_1 = 0.9$，$\beta_2 = 0.999$，最大训练轮数 150，使用早停的训练策略，当评价指标经过五轮训练没有明显提升时停止训练。

## 3.4 评价指标

为了对模型生成的推荐物品排名列表进行评估，本工作使用两种常见的 Top-K 推荐评价指标，命中率(Hit Ratio，HR)和归一化折损累计增益 (Normalized Discounted Cumulative Gain，NDCG)。HR@K 计算推荐物品列表排名前 K 个物品中包含用户下一个交互的物品的概率。NDCG@K 还考虑到物品在推荐列表中的排名因素，若用户下一个交互的物品在列表中排名越靠前则 NDCG@K 的得分就越高。本工作中选 K=5，10，多方面对推荐性能进行评测。

## 3.5 推荐效果

PreBERT 模型和蒸馏后得到 Mini-BERT 模型以及其他对比方法在三个数据集上的测试推荐结果如表 4 所示。

实验结果表明，本文将在文本数据集上经过预训练的 BERT 用于序列推荐能极大地提升推荐精度，并且蒸馏后得到的 Mini-BERT 模型相比于原始 BERT4Rec 模型也取得了一定的提升。第一阶段通过对预训练的 BERT 进行微调得到的 PreBERT 模型在所有数据集和评测指标下都显著优于其他对比方法。其中，在 ML-1M 数据集上取得了最佳也最稳定的推荐效果，与第二名

GRU4Re 相比，在 HR@5 和 NDCG@5 指标上分别提升了 14.35% 和 12.09%，在 HR@10 和 NDCG@10 指标上分别提升 17.15% 和 14.41%，说明 PreBERT 能够更准确地捕捉用户的兴趣，提高了推荐系统的准确性和覆盖范围。这一实验结果再一次展现了预训练的 BERT 具有一定的通用性知识以及强大的序列建模能力和适应能力。相比于 ML-1M 和 YooChoose 数据集，PreBERT 在数据稀疏程度较高的 Steam 数据集上提升的效果最不明显，体现了数据稀疏性问题为推荐系统领域引入复杂模型带来的挑战。

第二阶段对微调得到的 PreBERT 模型进行蒸馏，得到 Mini-BERT 模型。可以看到，Mini-BERT 在所有数据集和评测指标下也都优于其他对比模型。其中，在 ML-1M 数据集上 HR@10 和 NDCG@10 的提升最为显著，分别提升了 5.62% 和 3.81%，但同时在 HR@5 和 NDCG@5 上的提升并不显著，只有 0.93% 和

0.58%，说明 Mini-BERT 模型在更宽泛的推荐列表中表现得更好，能更好地识别用户的广泛兴趣，但在精准预测用户兴趣上还存在不足。值得一提的是，在 PreBERT 推荐性能提升最小的 Steam 数据集上，Mini-BERT 模型在评价指标上仍然取得了平均 2.64% 的提升效果，相比于 PreBERT 平均提升的 4.74% 只降低了 2.1%，说明本工作提出的算法在一定程度上能缓解数据稀疏性带来的推荐效果不佳的问题。

尽管 Mini-BERT 在一定程度上相比于其他对比方法有提升，但与蒸馏前的 PreBERT 模型相比，其推荐效果下降显著。这可能与本工作只采用了基于输出概率的蒸馏方法有关，未来的工作中可以考虑使用特征蒸馏、关系蒸馏等方法，以提升蒸馏后模型的推荐性能，避免性能显著下降的问题。

表 4 不同方法的推荐效果对比

| 数据集 | 评价指标 | BPR-MF | GRU4Rec | Caser | SASRec | BERT4Rec | PreBERT | 提升 | Mini-BERT | 提升 |
|---|---|---|---|---|---|---|---|---|---|---|
| ML-1M | HR@5 | 0.0091 | 0.0857 | 0.0268 | 0.0546 | 0.0793 | **0.0980** | 14.35% | <u>0.0865</u> | 0.93% |
| | NDCG@5 | 0.0062 | 0.0521 | 0.0136 | 0.0309 | 0.0489 | **0.0584** | 12.09% | <u>0.0524</u> | 0.58% |
| | HR@10 | 0.0184 | 0.1440 | 0.0467 | 0.0977 | 0.1338 | **0.1687** | 17.15% | <u>0.1521</u> | 5.62% |
| | NDCG@10 | 0.0091 | 0.0708 | 0.0200 | 0.0448 | 0.0665 | **0.0810** | 14.41% | <u>0.0735</u> | 3.81% |
| Steam | HR@5 | 0.0346 | 0.0695 | 0.0564 | 0.0609 | 0.0739 | **0.0775** | 4.87% | <u>0.0758</u> | 2.57% |
| | NDCG@5 | 0.0224 | 0.0471 | 0.0387 | 0.0419 | 0.0490 | **0.0513** | 4.69% | <u>0.0502</u> | 2.45% |
| | HR@10 | 0.0533 | 0.1079 | 0.0884 | 0.0952 | 0.1167 | **0.1221** | 4.63% | <u>0.1200</u> | 2.83% |
| | NDCG@10 | 0.0284 | 0.0595 | 0.0490 | 0.0529 | 0.0627 | **0.0657** | 4.78% | <u>0.0644</u> | 2.71% |
| YooChoose | HR@5 | 0.2193 | 0.2436 | 0.0868 | 0.2300 | 0.2771 | **0.3210** | 15.84% | <u>0.2831</u> | 2.18% |
| | NDCG@5 | 0.1364 | 0.1566 | 0.0593 | 0.1461 | 0.1813 | **0.2159** | 19.08% | <u>0.1843</u> | 1.65% |
| | HR@10 | 0.3392 | 0.3670 | 0.1267 | 0.3529 | 0.4021 | **0.4369** | 8.65% | <u>0.4127</u> | 2.63% |
| | NDCG@10 | 0.1751 | 0.1965 | 0.0721 | 0.1860 | 0.2218 | **0.2534** | 14.25% | <u>0.2262</u> | 1.98% |

注：加粗为一行中的最优得分，下划线为一行中的次优得分，提升指相比于最优对比方法的提升

## 3.6 随机映射对模型的影响

本文在数据预处理阶段将物品 ID 随机映射到 BERT 的 Token ID 上，以输入用户的历史行为序列。为了验证这种随机映射对模型的稳定性的影响，是否会导致模型性能的显著变化，本文在 ML-1M 和 Steam 数据集上进行了三次独立的

随机映射实验，得到的实验结果如表 5 所示。

对各组实验得到的评测指标计算标准差，可以看到，不管是在 ML-1M 数据集还是 Steam 数据集上，各评价指标的标准差都较小，最大的标准差也只有 $1.63 \times 10^{-3}$，最小的标准差为 $2.62 \times 10^{-4}$，表现出了极高的稳定性。这表明随机映射的方案是可行的，并不会对模型的性能产

生显著的影响。

#### 表 5 随机映射探究模型稳定性

| 数据集 | 实验 | HR@5 | NDCG@5 | HR@10 | NDCG@10 |
|--------|------|------|--------|-------|---------|
| ML-1M | 实验 1 | 0.0865 | 0.0524 | 0.01521 | 0.0735 |
| | 实验 2 | 0.0842 | 0.0515 | 0.1506 | 0.0729 |
| | 实验 3 | 0.0882 | 0.0548 | 0.1484 | 0.0742 |
| | 均值 | 0.0863 | 0.0529 | 0.1504 | 0.0735 |
| | 标准差<br>(×10⁻³) | 1.63 | 1.39 | 1.52 | 5.31 |
| Steam | 实验 1 | 0.0758 | 0.0502 | 0.1200 | 0.0644 |
| | 实验 2 | 0.0757 | 0.0501 | 0.1189 | 0.0639 |
| | 实验 3 | 0.0745 | 0.0496 | 0.1181 | 0.0635 |
| | 均值 | 0.0753 | 0.0500 | 0.1190 | 0.0639 |
| | 标准差<br>(×10⁻⁴) | 5.91 | 2.62 | 7.79 | 3.68 |

### 3.7 预训练参数的影响

为了探究初始化参数权重以及模型大小对性能的影响，本文设计了以下实验：将 GRU4Rec、SASRec、BERT4Rec 改为 12 层基本网络单元的堆叠，并且隐藏层维度设置为 768，多头注意力个数设置为 12，BERT 模型结构保持不变，但分别随机初始化整个模型参数、嵌入层参数、各层 Transformer 编码器中的参数进行实验。由于模型体量增大，为了使模型训练稳定，学习率均设置为 2e-5，训练 100 轮。在 ML-1M 数据集上进行实验得到的实验结果如表 6 所示。

#### 表 6 初始化参数影响

| 模型 | HR@5 | NDCG@5 | HR@10 | NDCG@10 |
|------|------|--------|-------|---------|
| GRU4Rec | 0.0174 | 0.0102 | 0.0366 | 0.0164 |
| SASRec | 0.0186 | 0.0108 | 0.0275 | 0.0137 |
| BERT4Rec | 0.0658 | 0.0399 | 0.0578 | 0.0434 |
| Scratch-BERT | 0.0111 | 0.0074 | 0.0193 | 0.0101 |
| Scratch-Embed | 0.0126 | 0.0070 | 0.0258 | 0.0113 |
| Scratch-Layer | 0.0140 | 0.0086 | 0.0225 | 0.0114 |
| PreBERT | 0.0980 | 0.0584 | 0.1687 | 0.0810 |

注：Scratch-BERT 指将整个 BERT 模型进行随机初始化参数，Scratch-Embed 指对 BERT 的嵌入层进行随机初始化，Scratch-Layer 指对 BERT 的中间层进行随机初始化。

与表 3 中的数据对比可知，增大 GRU4Rec、SASRec 和 BERT4Rec 的网络层数以及隐藏维度并没有提升模型性能，反而导致了推荐效果的显著下降。由此可以看出，如果只是简单地扩展模型结构，由于推荐数据集的数据稀疏性问题的存在，没有足够的数据支持复杂模型的稳定训练。对 BERT 参数的随机初始化实验结果也同样说明了这一问题，三种初始化方式均导致了模型性能的显著下降。这表明在文本数据上经过预训练的 BERT 参数权重具有一定的通用性，当迁移到其他领域时也可以使模型稳定收敛。不仅如此，预训练 BERT 模型还表现出了出色的推荐性能。

### 3.8 超参敏感分析
#### 3.8.1 遮掩比率对 PreBERT 的影响

由于 PreBERT 对序列中被遮蔽的物品进行预测，因此遮掩比率 $\rho$ 直接影响损失函数，是影响模型性能的一个关键因素。为了探究不同的遮掩比率对 PreBERT 推荐性能的影响，实验中固定其他参数不变，将遮掩比率 $\rho$ 从 0.15 取值到 0.85，在上述三个数据集上进行实验，得到的实验结果如图 4 所示。

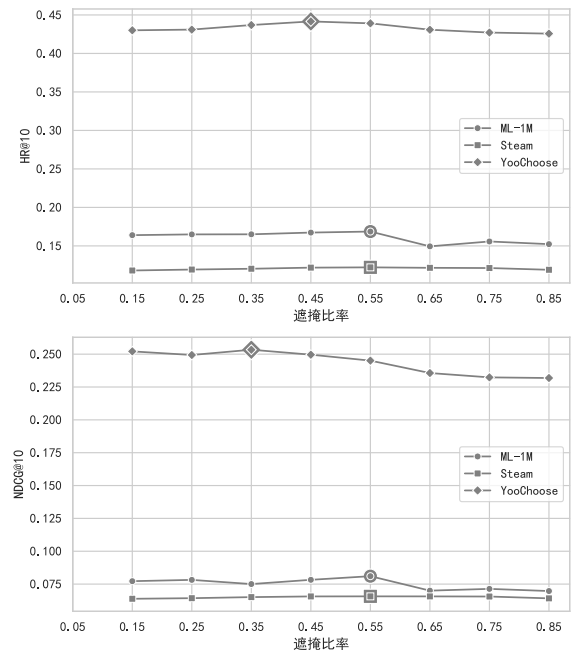

图 4 不同的遮掩比率对 PreBERT 推荐效果的影响

由图可以看出，当遮掩比率较小时，随着遮掩比率的增大，PreBERT 的推荐性能提高，例如 $\rho = 0.35$ 时的 HR@10、NDCG@10 指标值均高于 $\rho = 0.15$ 时；但当时 $\rho > 0.55$ 时，随着遮掩比率的增大，模型的性能开始下降。这一实验结果说明

遮掩比率过小或过大都将对模型性能产生影响，遮掩比率过小，模型没有足够的机会从遮蔽的物品中进行学习，无法充分对模型进行训练；遮掩比率过大会产生较多的物品需要模型进行预测，学习难度增加，上下物品序列信息的丢失会导致模型推荐性能下降。

### 3.8.2 损失函数系数对蒸馏的影响

如前文所述，蒸馏时使用的损失函数为式 (10)。其中 $\alpha$ 越大，硬目标的影响越大，软目标的影响越小，即受真实物品值的影响越大而受教师网络输出的物品概率分布的影响越小。

为了探究对蒸馏后所得到的 Mini-BERT 模型性能的影响，将 $\alpha$ 取值为 $[0, 0.3, 0.5, 0.7, 1]$ 进行实验，得到的实验结果如图 5 所示。

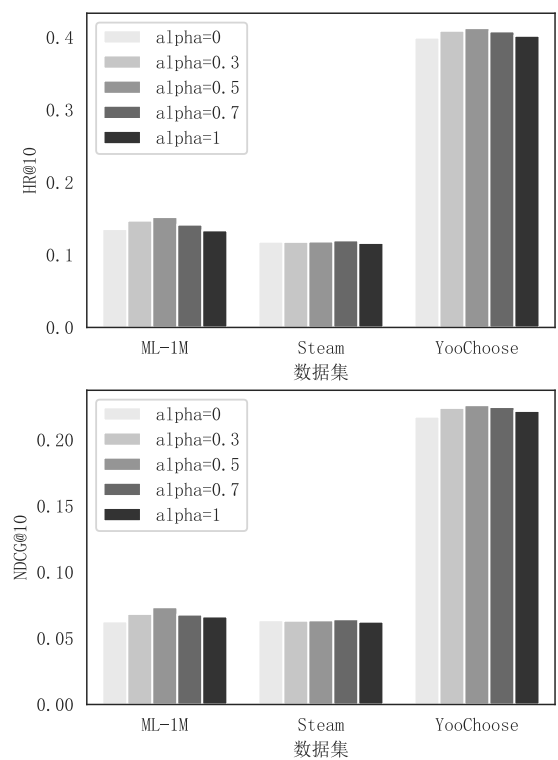

图 5 不同蒸馏系数对 Mini-BERT 性能的影响

当 $\alpha = 0$ 时，$\mathcal{L} = T^2 \mathcal{L}_{soft}$，这意味着蒸馏过程完全依赖于软目标的影响；当 $\alpha = 1$ 时，$\mathcal{L} = \mathcal{L}_{hard}$，完全只受硬目标影响。由实验结果可知，当损失函数只受单一目标影响时，蒸馏的效果在所有数据集上均不如将两种目标同时结合起来的效果。在 ML-1M 和 YooChoose 数据集上，当 $\alpha = 0.5$ 时，蒸馏效果最佳。而在 Steam 数据集上，$\alpha$ 的改变

随模型性能的影响并不显著，模型性能在 $\alpha = 0.7$ 时达到最佳。

### 3.8.2 温度对蒸馏的影响

如前文所述，温度对蒸馏的影响主要在于对教师网络 PreBERT 输出概率分布的影响，温度越高输出的概率分布越平滑，温度越低输出的概率分布的确定性越强。

为了探究温度对蒸馏的具体影响，本文将温度 $T$ 从 1.1 取值到 1.7，在 ML-1M 数据集上进行实验，得到的实验结果如图 6 所示。

由实验结果可以看出温度过高过低都会对模型的性能造成影响，适当的升温可以提升模型的性能，当温度过高超过一定界限，例如在温度 $T > 1.5$ 时，升高温度会导致模型性能的剧烈下降。

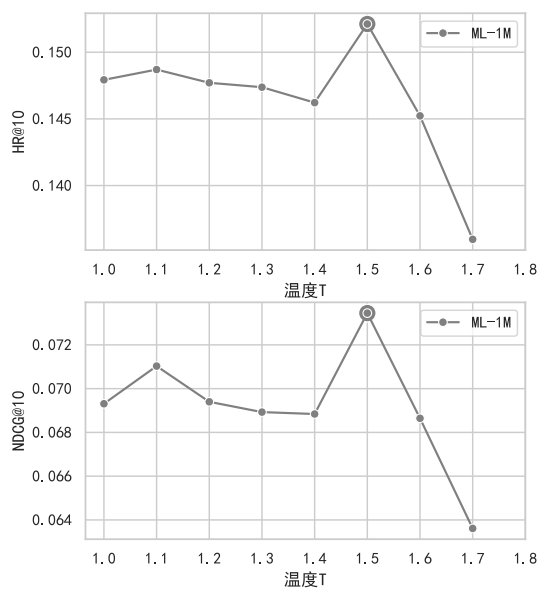

图 6 温度对蒸馏的影响

## 4 总结

本文为了克服数据稀疏性和实时性需求给预训练语言模型在推荐系统中的应用带来的难题，提出了一种基于预训练语言模型和知识蒸馏的轻量化序列推荐算法，第一阶段使用预训练语言模型在推荐数据集上进行微调训练，第二阶段对已经训练好的语言模型进行知识蒸馏，得到一个轻量级的模型。实验结果证明了本方法的可行性，通过本文算法可以将预训练语言模型的先验

知识跨领域迁移到推荐领域中，经过知识蒸馏精炼萃取得到的小型模型也能在提升推荐性能的同时又能提供快速推理服务，满足推荐系统实时性的需求，实现轻量化推荐。但本文只尝试了使用知识蒸馏对大型语言模型进行压缩，未来如何将有着高推荐精度的大型语言模型应用于实际推荐场景中仍值得进一步探究。

## 参考文献

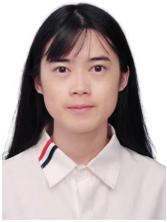
李莉（2002—），学士，主要研究领域为推荐系统。
E-mail: lili0516@mail.ustc.edu.cn

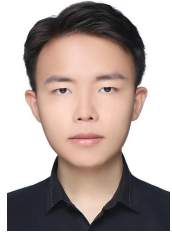
刘芷町(1998—)，博士研究生，主要研究领域为时间序列、推荐系统。
E-mail: zhiding@mail.ustc.edu.cn

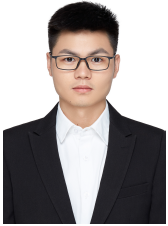
程明月（1993—），通信作者，博士，特任副研究员，主要研究领域为时间序列分析、推荐系统。
Email: mycheng@ustc.edu.cn